\documentclass[12pt]{article}
\usepackage[utf8]{inputenc}
\usepackage{fullpage}
\usepackage{enumerate}
\usepackage{hyperref}
\usepackage{xcolor}
\usepackage{wordlike}
\usepackage{tcolorbox}
\usepackage{colortbl}

\newcommand{\note}[1]{\textcolor{red}{#1}}
\newcommand{\noop}[1]{}

\usepackage{fancyhdr,lastpage}
\title{Knowledge Scientists \\ \Large Unlocking the data-driven organization}

\author{George Fletcher\thanks{Eindhoven University of Technology (\href{mailto:g.h.l.fletcher@tue.nl}{\footnotesize{{\tt g.h.l.fletcher@tue.nl}}})}, 
Paul Groth\thanks{University of Amsterdam
 (\href{mailto:p.groth@uva.nl}{\footnotesize{{\tt p.groth@uva.nl}}})}, 
and
Juan Sequeda\thanks{data.world 
 (\href{mailto:juan@data.world}{\footnotesize{{\tt juan@data.world}}})}}

\date{April 2020}

\begin{document}

\maketitle

\noop{
\begin{tcolorbox}
\noindent {\em Do you do Knowledge Science?}

\noindent Do you...
\begin{itemize}
\item lead conversations with your organization’s stakeholders to understand their pain points? \note{gf: data pain points?}

\item debate with stakeholders about the definition of a ''customer''?
\item draw whiteboard sketches that define the schemas and models for your team's data?
\item meet people who come to you asking about data?
\item find yourself in the middle of conversations between data producers and consumers?
\item maintain a data catalog?
\item negotiate with data owners?
\item wrangle and clean data?
\item source data for machine learning and applications?
\item feel responsible for your team's data?
\end{itemize}
\end{tcolorbox}
}

\section*{The Data Driven Organization - the new secular trend}

The rising importance of data in organizations over the last two decades is undeniable.   Organizations across all sectors are increasingly undergoing deep transformation and restructuring towards data-driven operations.  Looking back, we can identify two earlier trends which brought us to the data-driven organization. 

In the 2000’s with the rise of the Web, and driven by Moore’s Law, organizations of all sizes could, for the first time, easily capture and process massive data collections using commodity distributed computing platforms such as Hadoop \cite{DBLP:conf/osdi/DeanG04} and NoSQL databases \cite{DBLP:conf/osdi/ChangDGHWBCFG06,DBLP:conf/sosp/DeCandiaHJKLPSVV07}.  
Big Data became democratized.  

In the 2010’s, with the rise of Big Data and dropping prices in GPU’s, organizations could increasingly perform sophisticated data analytics and extract value from their data using commodity machine learning solutions such as TensorFlow \cite{DBLP:conf/osdi/AbadiBCCDDDGIIK16}.  
Data analytics and the AI revolution became democratized.  

The Big Data trend placed increasing importance for organizations on being able to collect and harness data.   
As this trend matured, organizations with competitive advantage are the ones which identified and developed the role of Data Engineer to manipulate large amounts of data \cite{anderson} and Data Stewards to manage data governance and workflows \cite{datasteward2018}.  
The AI revolution has lead to the vital need for organizations to be able to draw value from their data \cite{NAP18374,HalevyNP09}.  
This lead to the rise of the Data Scientist as a critical role \cite{DataScientistHBR}.  

The AI revolution and the rise of cloud computing make possible the shift to the data-driven organization.  In the data-driven organization, the central role of data highlights the need for reliable and clean data.  If you don’t have clean and reliable data, your AI, machine learning and analytics are worthless: garbage in, garbage out.  Unreliable, erroneous, and incomplete data lead to critical bottlenecks in processing pipelines and, ultimately, service failures, which are disastrous for the competitive performance of the organization.   

Currently, the responsibility for making data reliable is implicitly shared between the data engineer, data stewards and data scientist roles.  Given its central importance, those organizations which recognize and react to the need for reliable data will have the advantage in the coming decade.

We argue that the technologies for reliable, clean, meaningful, beautiful data are driven by distinct concerns and expertise which complement those of the data scientist and the data engineer.  Those organizations which identify the central importance of meaningful, explainable, reproducible, and maintainable data will be at the forefront of the democratization of reliable data.   We call the new role which must be developed to fill this critical need the {\bf Knowledge Scientist}.

\begin{table}
\caption{Key secular trends in recent data history.}\label{table:trends}
\begin{center}
\begin{tabular}{|p{2.5cm}|p{2.5cm}|p{3cm}|p{4cm}|p{2.5cm}|}
\hline
\hline
\rowcolor{gray}{\bf Context} & {\bf Trend} & {\bf Organizational Need}& {\bf Technology}& {\bf Role}\\
\hline
\hline
Web + Moore's Law & Big Data  & Harness and collect data & 
Commodity distributed computing platforms (e.g., Hadoop) & Data Engineer \\
\hline
Big Data + GPU Compute & AI Revolution& Draw value from data &
Commodity machine learning (e.g., TensorFlow, SciPy) & Data Scientist\\ 
\hline
AI Revolution + Cloud Computing &
Data-Driven Organization, Digital Transformation &
Rely on data  &
Clean, meaningful, beautiful data technologies (e.g. knowledge graphs, data wrangling systems, data catalog platforms) &
{\bf \textcolor{blue}{Knowledge Scientist}}\\
\hline
\hline
\end{tabular}
\end{center}
\end{table}

\section*{Unpacking ``rely on data''}

The organizational shift we articulated above means that the products of data science are no longer just a ``bonus'' but are central to an organization's value. This means that organizations must be able to {\bf rely} on the data. But what does this mean in practice? 

Consider the following example in an e-commerce company: The business user is trying to answer the question “how many orders were placed in a given time period per their status?” As simple as this question may seem, there is a lot to be unpacked. 

When analyzing the data, are there missing values, and if so, what should they be? 
What should the reference time zone be? 
Which currency unit should we use? 
What’s the assumed exchange rate? 
{\bf (Property 1)}

Putting the data aside, what exactly is the definition of ``an order''? 
Is it when a customer clicked ``place order'' on the website? 
Or is it when the money from the customer has been received? 
Or is it when the package has been delivered to the customer? 
Each of these are valid definitions of ``an order''. {\bf (Property 2)}

What is the correct source of the data? Is it data coming from the order management system? Or perhaps the accounting system? {\bf (Property 3)} 

Can the data be easily provided to business analysts in the tools of their choice and can it be easily fed into churn models? {\bf (Property 4)} 

Finally, is the business leader confident that the results given are truly up-to-date and will be reproducible in future analyses? {\bf (Property 5)}.

Being able to systematically answer these kinds of questions is fundamental for having reliable data. We summarize these questions as a series of properties that can be checked in Table \ref{table:properties}. 

\begin{table}
\caption{Properties of reliable data.}\label{table:properties}
\begin{tcolorbox}
\begin{enumerate}
\item {\em Reliable data is clean data.} An obvious place to start is traditional notions of clean data: that it is uniform (e.g. timestamps are all the same), that it’s valid in conforming to business rules and well defined schemas, that it is complete, that it accurately reflects reality, that it’s consistent with other data in the organization. 
\item  {\em Reliable data is grounded in shared meaning spaces.} What does a column mean? Can I understand that without talking to the initial producer(s) of the data? Is the meaning adopted across the organization and do we have a shared understanding? Can one tie the data easily to human understandable definitions? 
\item  {\em Reliable data is data in context.} Clean data with shared meaning is not enough. At first blush, data may look pristine but if we don’t know where it comes from (its lineage), how it was sourced, and if we have the rights to use it, it can become a massive liability. What if we didn’t obtain consent for its usage and violate regulatory laws? What if the licenses under which the data is obtained constrain the usage data and we are now financially liable? What if we use data where it was cleaned appropriately for one task but will fail for another?   We can rely on data when we know the organizational context in which it’s situated, developed and sourced. 
\item  {\em Reliable data is data accessible in a standardized format.} Can data be easily imported in existing tools? Are the tools and programs available? Is it using open formats? 
\item  {\em Reliable data is maintained.} An organization can only rely on data that reflects the current state of an organization's world and that is kept up-to-date and reproducible for future analyses.
\end{enumerate}
\end{tcolorbox}
\end{table}

All these properties require deeper knowledge and understanding of the data. Essentially, we can rely on data only when we acknowledge the fact that we cannot ignore the organizational activities within which it is situated. Recent work on understanding data science practices \cite{muller2019data, workshopChiDatascience2019, borgmanbook} highlights the importance of such an acknowledgement. 

Data wrangling is often said to be 80\% of the work of data science \cite{ruiz_2017}.  
Typically this is seen as boring annoying grunt work - data janitorial work that people don't want to do and get stuck with it \cite{lohr_2014}. 
{\em However, the problem is not about cleaning data by eliminating white spaces and replacing wrong characters.  It’s about understanding the ecosystem between people, data and tasks in an organization and about communicating, documenting, and maintaining that knowledge.} 
This is why it takes 80\% of the effort.  
Our contention is that this is vital knowledge work at the center of any data-driven organization.

In typical organizations, the knowledge work to create reliable data is ad-hoc and the results and practices are not shared \cite{DBLP:journals/debu/StonebrakerI18}. 
Furthermore, in data science teams, a data scientist or data engineer might do this knowledge work, but is not equipped, trained or incentivized to do so. 
Indeed, from our experience, the knowledge work (e.g. 8 hour conference calls, discussions, documentation, long slack chats, confluence spelunking ) required to create reliable data is often not valued by managers or employees themselves. 
The tasks and functions of creating reliable data are never fully articulated and thus responsibility is diffuse or non-existent. 
Who should be responsible?

\section*{The role of the Knowledge Scientist}
The Knowledge Scientist is responsible. 

\paragraph*{Who is a knowledge scientist? }

The Knowledge Scientist is the person who builds bridges between data and business requirements, questions, and needs. 
Their goal is to document knowledge by gathering information from Business Users, Data Scientists, Data Engineers and their environment with the goal of making reliable data that can then be used effectively in a data driven organization.

Returning to our scenario of order placement over time. 
The knowledge scientist gets in a room with business users and starts the discussion about the business question. 
During this discussion, it becomes apparent that multiple business users use the same word to mean different things. 
After further discussion, there may be an agreement that the correct use of the term ``order'' is when money has been received and the package has been delivered. 
With this information, the knowledge scientist can work with the data engineer to identify that the Order management system should be used and with data scientists to figure out how best to enable churn modeling. 
Furthermore during analysis of the data, different time zones are being recorded. 
The knowledge scientist would then go back and have further discussions with the business users to define what would constitute an agreed meaning for a time period. 
Given that the HQ of the company is on the west coast, they all agree that PST is the reference timezone. 
We can keep going on with this example, but what is important to note that the knowledge scientist is the person that is driving the discussions between the business users, data scientists, and data engineers, documenting the decisions and defining how the data corresponds to the business meaning. 
At the end, the knowledge scientist is able to generate reliable data that can then be passed on to the data scientist for analysis. 

\paragraph*{What skills does a knowledge scientist need to have?}

Knowledge science work is technical work. 
Knowledge Scientists use skills and techniques such as data modeling, data integration, knowledge representation, ontology engineering to manifest what they learned from business users. 
The output is a data model that represents how the business user sees the world. 
They can align this data model with other models derived from talking to other business users. 
Furthermore, while working with Data Engineers, the knowledge scientist is fluent in data access and transformation methods such as query and programming languages. 
They can transform the data being provided by the data engineer and map it to the business meaning provided by the business user. 
They are conversant in analytical and machine learning methods.

Knowledge science work is people work.  
The knowledge scientist has excellent communication skills that can be applied to both the business user and data engineer. 
The knowledge scientist is both a ``people person'' and a ``geek'', who is comfortable with the context-dependent, dynamic, and collaborative nature of meaning making from data.

\paragraph*{Haven’t we seen this before?}

Knowledge science has its roots in the knowledge engineering approaches of the 1980s and 1990s \cite{DBLP:journals/dke/StuderBF98}. 
In that world, skills such as knowledge identification, knowledge elicitation, and knowledge specification were taught and used.  
These are lost arts in industry today and particularly in the data science context. 
We believe that revisiting these approaches will be a key part of developing both the instructional curriculum and also tooling needed to support the knowledge scientist.

\section*{Implications}
 
We need to empower people and organizations to produce reliable data.   This requires rethinking across the board: organizational structures, academic training, and advances in the study of knowledge. The following are some of our initial thoughts on what this means for these various actors.

For organizations, introduce the role of a knowledge scientist into your organization. A simple step is to seek out existing team members who play this role. They might be business analysts, data scientists, product owners, data stewards or data engineers. We have even seen sales representatives who have played this role. By acknowledging this role, you can elevate the importance of reliable data in your organization and focus on the development of the skills outlined above. Another step is to communicate your experience with knowledge scientists: What do they bring to the organization? What infrastructure do they need? How have they impacted your data estate? Only through this communication can we effectively find the practices that empower knowledge scientists. For tool developers, there is paucity of tools that help knowledge scientists - now might be a time to see what you can do. 

For educators, new courses and content are needed that is {\em integrative}. A knowledge scientist does not just require the skills of data engineering, machine learning, knowledge acquisition, communications, and human computer interaction. They require pieces of all these. To do so, one approach would be to provide {\em pathways} through the excellent content available for these various disciplines. It is also perhaps time to revisit classic knowledge engineering and data modeling knowledge and update it for this new area. Finally, as we have already noted, knowledge scientists exist. We see a tremendous opportunity to learn from this practical experience. 

For researchers, again our call is to be integrative. We need to study the tripartite relationship between data/knowledge models, their corresponding query languages, and the people both using and producing reliable data. We need to understand how people perceive the way data is modeled and represented and its organizational embedding. In order to do so, we need to work with scientists and experts across communities to design methodologies, experiments and user studies. This requires bringing together data management expertise in theory, systems and semantics with communities who study people (e.g. human data interaction) and those who are actively using data (e.g. data journalist, political scientist, life science, etc.) In this integrative approach what kinds of questions might be interesting to ask? Here's a few to get going:  What is the role and function of data and knowledge modeling in organizations? Why do we keep inventing new data models? What are the affordances necessary to help people in creating and using data models? Are new organizational roles needed for data and knowledge modeling and management? 

The organizational structures, tools, methodologies and techniques to support and make possible the work of knowledge scientists are still in their infancy. Let's make their work easier.

\section*{Unlocking the data-driven organization}

We have argued that there is a fundamental role  -- the knowledge scientist -- that has been overlooked in data driven organizations. As organizations not only use data but increasingly {\em rely} on data, it is time to empower the people who are central to this transformation. 

\bibliographystyle{unsrturl}
\bibliography{bibliography}

\end{document}